\documentclass[12pt,draftcls,onecolumn]{IEEEtran}
\usepackage{amsmath}
\usepackage{amssymb}
\usepackage{graphicx}
\usepackage{url}
\usepackage{color}

\makeatletter

\newcommand{\CC}{\mathbb{C}}

\newcommand{\RR}{\mathbb{R}}
\newcommand{\ZZ}{\mathbb{Z}}

\newcommand{\cv}{\mathbf{c}}
\newcommand{\hv}{\mathbf{h}}

\newcommand{\rv}{\mathbf{r}}

\newcommand{\tv}{\mathbf{t}}
\newcommand{\uv}{\mathbf{u}}
\newcommand{\vv}{\mathbf{v}}

\newcommand{\xv}{\mathbf{x}}
\newcommand{\yv}{\mathbf{y}}
\newcommand{\zv}{\mathbf{z}}

\newcommand{\Vc}{\mathcal{V}}

\newcommand{\vol}{{\rm vol}}

\newcommand{\diag}{\mathsf{diag}}

\newcommand{\vect}{\mathsf{vec}}
\newcommand{\Tr}{{\rm Tr}}

\newtheorem{prop}{Proposition}
\newtheorem{remark}{Remark}

\newtheorem{lem}{Lemma}
\newtheorem{cor}{Corollary}

\makeatother

\title{An Error Probability Approach \\ to MIMO Wiretap Channels}
\author{Jean-Claude Belfiore and Fr\'ed\'erique Oggier
\thanks{Jean-Claude Belfiore is with Telecom ParisTech, CNRS, UMR 5141, France.
Fr\'ed\'erique Oggier is with Division of Mathematical Sciences, School of
Physical and Mathematical Sciences, Nanyang Technological University,
Singapore. Email:belfiore@telecom-paristech.fr, frederique@ntu.edu.sg.
Part of this work appeared as an invited paper in ICC 2011 \cite{icc}.
}}
\begin{document}
\maketitle
\begin{abstract}
We consider MIMO (Multiple Input Multiple Output) wiretap channels, where a legitimate transmitter Alice is communicating with a legitimate 
receiver Bob in the presence of an eavesdropper Eve, and communication is done via MIMO channels. 
We suppose that Alice's strategy is to use a codebook which has a lattice structure, which then allows her 
to perform coset encoding. We analyze Eve's probability of correctly decoding the message Alice meant to Bob, 
and from minimizing this probability, we derive a code design criterion for MIMO lattice wiretap codes.
The case of block fading channels is treated similarly, and fast fading channels are derived as a particular case. 
The Alamouti code is carefully studied as an illustration of the analysis provided.
\end{abstract}


\section{Introduction}

Wiretap channels were introduced by Wyner \cite{W-75} in the seventies as broadcast channels, 
where a legitimate transmitter Alice communicates with a legitimate receiver Bob through a noisy 
communication channel in the presence of an eavesdropper Eve. They have attracted a regain 
of interest recently, in particular in the context of physical layer security. 
We consider MIMO (Multiple Input Multiple Output) wiretap channels, for which the secrecy capacity, 
that is the maximum amount of information that Alice can transmit confidentially to Bob, is known  
\cite{yaKW-10,OH,LS-08}. We consider an alternative approach, which consists of studying the probability 
that Eve correctly decodes the message meant to Bob, as initiated in \cite{sec-gain,OSB} for Gaussian channels. 
An early work by Hero \cite{hero} proposed a non-information theoretical approach to secrecy in MIMO channels, 
where a code design was proposed, based on the assumption that Eve is doing a non-coherent decoding. 
In \cite{Wong}, the model of wiretap channel is further used to study secret sharing over fast fading MIMO channels.

We consider the case where Alice transmits lattice codes using coset encoding, which requires two nested 
lattices $\Lambda_e \subset \Lambda_b$, and Alice encodes her data in the coset representatives of $\Lambda_b/\Lambda_e$. 
Both Bob and Eve try to decode using coset decoding. It was shown in \cite{sec-gain} for Gaussian channels that a wiretap 
coding strategy is to design $\Lambda_b$ for Bob (since Alice knows Bob's channel, she can ensure he will decode with high probability), while $\Lambda_e$ is chosen to maximize Eve's confusion, characterized by a lattice invariant called 
secrecy gain, under the assumption that Eve's noise is worse than the one experienced by Bob. 
The contribution of this work is to generalize this approach to MIMO channels (and in fact block and fast fading channels as particular cases). We compute Eve's probability of making a correct decoding decision, and deduce how the lattice \textcolor{black}{$\Lambda_e$} should be designed to minimize this probability.
\textcolor{black}{A MIMO wiretap channel will then consist of two nested lattices $\Lambda_e\subset \Lambda_b$ where $\Lambda_b$ is designed to ensure Bob's reliability, while $\Lambda_e$ is a subset of $\Lambda_b$ chosen to increase Eve's confusion.}
\textcolor{black}{
More precisely, we prove that to minimize Eve's average probability of correct decoding, 
a code design is 
\[
\min_{\Lambda_e}\sum_{\xv\in\Lambda_e\backslash\{0\}}\frac{1}{\det(XX^*)^{n_e+T}}
\]
where $n_e$ is Eve's number of antennas, $T$ is the coherence time of the MIMO channel, and $\xv$ is the vectorized codeword $X$. As a corollary, we derive a design criterion for a block fading channels where all numbers of antennas are the same, namely
\[
\min_{\Lambda_e}\sum_{\xv\in\Lambda_e\backslash\{0\}}\frac{1}{(\prod_{i=1}^n||\xv_i||^2)^{1+T}}
\]
which in turn gives a criterion for a fast fading channel:
\[
\min_{\Lambda_e}\sum_{\xv\in\Lambda_e\backslash\{0\}}\frac{1}{(\prod_{i=1}^n|x_i|^2)^{2}}.
\]
}

This paper is organized as follows: in Section \ref{sec:gauss}, we recall how Eve's probability of correct decision is 
derived for Gaussian channels, and extend the computation to include the case where low dimensional lattice codes are used. 
Section \ref{sec:mimo} is the chore part of this paper, which contains Eve's probability of correctly decoding the confidential 
message when her channel from Alice is a MIMO channel. We consequently treat the case of block and fast fading channels 
in Section \ref{sec:bffading}. The relevance of our approach is illustrated in Section \ref{sec:alamouti} where the Alamouti code is studied following the newly introduced techniques.

%
%
%
\section{Gaussian channels}
\label{sec:gauss}

We first consider a Gaussian wiretap channel, modeled by
\begin{equation}\label{eq:wiretap}
\begin{array}{ccl}
\yv &= & \xv + \vv_b\\
\zv &= & \xv + \vv_e
\end{array}
\end{equation}
over $n$ complex channel uses, where $\xv\in\CC^n$ is the transmitted signal, $\vv_b\in\CC^n$ and $\vv_e\in\CC^n$ denote the Gaussian noise at Bob, respectively Eve's side, both with coefficients which are zero mean, and have respective variance $\sigma_b^2$ and $\sigma_e^2$, where $\sigma_{e}$ is assumed larger than $\sigma_{b}$.
We assume that Alice knows Bob's channel $\sigma_b$, and uses $\ZZ[i]-$lattice codes, namely
$\xv\in\Lambda$, where $\Lambda$ is an $m$-dimensional complex lattice\footnote{\textcolor{black}{Note that in the theoretical computer science literature, the dimension of a lattice is defined as the number of rows of $M$, whereas the rank of a lattice is defined as the number of columns of $M$.}}, which can be described by its generator matrix $M$ \cite{CS-98}:
\[
\Lambda = \{ \xv = M \uv  ~|~ \uv \in \ZZ[i]^m \},
\]
and the columns of $M$ form a linearly independent set of vectors in $\CC^n$ (so that $m\leq n$) which form a basis of the lattice.

Alice performs coset encoding \cite{W-75}: she chooses a lattice $\Lambda_b$ that she partitions into a union of disjoint cosets $\Lambda_e+\cv$, with $\Lambda_e$ a sublattice of $\Lambda_b$ and $\cv$ an $n$-dimensional vector which encodes her data.
Alice then randomly chooses a random vector $\rv\in\Lambda_e$ so that the transmitted lattice point $\xv\in\Lambda_b$ is finally
\begin{equation}\label{eq:xsent}
\xv = \rv + \cv \in \Lambda_e + \cv.
\end{equation}
Why coset encoding is actually beneficial for wiretap lattice codes is illustrated in \cite{OSB}.

Recall from \cite{sec-gain}\footnote{A real channel was considered in \cite{sec-gain}, the extension to the complex case discussed here is immediate.} that when $m=n$, the probability $P_c$ of correct decision when doing coset decoding is 
\textcolor{black}{
\begin{eqnarray*}
P_c &=& \frac{1}{(2\pi\sigma^2)^{n}}\sum_{\tv\in\Lambda_e}\int_{\Vc(\xv+\tv)}e^{-||\yv-\xv ||^2/2\sigma^2}d\yv
\\
&=&
\frac{1}{(2\pi\sigma^2)^{n}}\sum_{\tv\in\Lambda_e}\int_{\Vc(\Lambda_b)}e^{-||\uv+\tv ||^2/2\sigma^2}d\yv
\end{eqnarray*}
where $\Vc(\Lambda_{b})$ denotes the Voronoi region of $\Lambda_b$ and $\uv=\yv-\xv-\tv$. 
Equality holds because infinite lattice constellations are considered (this gives an upper bound on finite lattice constellations).
Since Bob's received vector $\yv$ is most likely to lie in the Voronoi region around the transmitted point, the terms corresponding to $\tv\neq 0$ are negligible, which yields the well known bound on the probability $P_{c,b}$ of Bob's correct decision:
\[
P_{c,b} \leq \frac{1}{(2\pi\sigma_{b}^2)^{n}}\int_{\Vc(\Lambda_{b})}
e^{-\Vert\uv\Vert^{2}/2\sigma_{b}^{2}}d\uv.
\]
Regarding the probability $P_{c,e}$ of Eve's correct decision in doing coset decoding, note that 
\[
\frac{1}{(2\pi\sigma_e^2)^{n}}\sum_{\tv\in\Lambda_e}\int_{\Vc(\Lambda_b)}e^{-||\uv+\tv ||^2/2\sigma_e^2}d\yv
=
\frac{1}{(2\pi\sigma_e^2)^{n}}\int_{\Vc(\Lambda_b)}\sum_{\tv\in\Lambda_e}e^{-||\uv+\tv ||^2/2\sigma_e^2}d\yv
\]
and since $\sum_{\tv\in\Lambda_e}e^{-||\uv+\tv ||^2/2\sigma_e^2}$ reaches its maximum when $\uv\in\Lambda_e$ (see Remark 2 in \cite{semantic}), 
we find that
\[
P_{c,e}\leq
\frac{1}{(2\pi\sigma^2)^{n}}\sum_{\tv\in\Lambda_e}\int_{\Vc(\Lambda_b)}e^{-||\tv ||^2/2\sigma^2}d\yv
=
\frac{\vol(\Lambda_{b})}{(2\pi\sigma_{e}^{2})^{n}}
\sum_{\tv\in\Lambda_{e}}e^{-\Vert\tv\Vert^{2}/2\sigma_{e}^{2}},
\]
where $\vol(\Lambda_b)$ is defined to be $\sqrt{\det(MM^*)}$.
}
We need to discuss the case where $m<n$ before proceeding. The notation we will use refers to Bob's channel, though the same holds for Eve's.

The decoding rule for a Gaussian channel (\ref{eq:wiretap}) when $m<n$ is similarly to the case $m=n$ given by
\[
\min_{\xv}||\yv-\xv||^2,
\]
where $\yv=\xv'+\vv_b$ is the noisy message at the receiver when $\xv'$ is sent, except that now, $\xv'=M\uv'$ and $\xv=M\uv$ where $M$ is an $n\times m$ complex matrix.
By performing a $QR$ decomposition of $M$, we get
\[
M = QR = Q
\begin{bmatrix}
R' \\ {\bf 0}\\
\end{bmatrix}
\]
with $R'$ an upper triangular $m\times m$ matrix, and $Q$ a unitary $n\times n$ matrix, \textcolor{black}{whose Hermitian transpose is denoted by $Q^*$}. Thus
\[
\min_{\xv}||\yv-\xv||^2=\min_{\uv}||Q^*(M\uv'+\vv_b)-Q^*M\uv||^2=\min_{\uv}|| (R\uv'+Q^*\vv_b)-R\uv  ||^2
\]
that is
\[
\min_{\uv\in\ZZ[i]^m}
\left\Vert
\begin{bmatrix}
R'\uv' \\ {\bf 0}\\
\end{bmatrix}
+
\vv_b'-
\begin{bmatrix}
R'\uv \\ {\bf 0}\\
\end{bmatrix}
\right\Vert^2
\]
where $\vv_b'$ is a new noise vector with the same noise statistics as $\vv_b$ since $Q^*$ is unitary.
It is now clear from the above minimization that
\[
\arg\min_{\uv} ||\yv-M\uv||^2 = \arg\min_{\uv} || R'\uv' + \vv_b''-R'\uv||^2
\]
where $\vv_b''$ is the noise vector $\vv_b'$ where the last $n-m$ rows have been ignored \textcolor{black}{($R'\uv'+\vv''$ is a vector containing the first $m$ elements of $Q^*\yv$)}, and thus the problem
of decoding an $m$-dimensional lattice in an $n$-dimensional space can be reduced to perform the decoding in an
$m$-dimensional space, showing that what matters is the dimension of the lattice, and not the one of the ambient space.
Consequently, we have
\begin{eqnarray}
P_{c,b} &=&\frac{1}{(2\pi\sigma_{b}^2)^{m}}\int_{\Vc(\Lambda_{b})}
e^{-\Vert\uv\Vert^{2}/2\sigma_{b}^{2}}d\uv,\label{eq:Pcb}\\
P_{c,e}&\leq&\frac{\vol(\Lambda_{b})}{(2\pi\sigma_{e}^{2})^{m}}
\sum_{\rv\in\Lambda_{e}}e^{-\Vert\rv\Vert^{2}/2\sigma_{e}^{2}}
\label{eq:Pce},
\end{eqnarray}
when Alice sends an $m$-dimensional lattice (living in an $n$-dimensional space) to Bob.
We are now ready to analyze the MIMO case.

%
%

\section{The MIMO case}
\label{sec:mimo}

We now consider the case when the channel between Alice and Bob, resp.
Eve, is a \textcolor{black}{quasi-static} MIMO channel with $n_{t}$ transmitting antennas at Alice's
end, $n_{b}$ resp. $n_{e}$ receiving antennas at Bob's, resp. Eve's
end, and a coherence time $T$, that is:
\begin{equation}
\begin{array}{ccl}
Y & = & H_{b}X+V_{b}\\
Z & = & H_{e}X+V_{e},
\end{array}\label{eq:MIMO}
\end{equation}
where the transmitted signal $X$ is a $n_{t}\times T$ matrix, the
two channel matrices are of dimension $n_{b}\times n_{t}$ for $H_{b}$
and $n_{e}\times n_{t}$ for $H_{e}$, and $V_{b}$, $V_{e}$ are
$n_{b}\times T$, resp. $n_{e}\times T$ matrices denoting the Gaussian
noise at Bob, respectively Eve's side, both with coefficients zero
mean, and respective variance $\sigma_{b}^{2}$ and $\sigma_{e}^{2}$.
The fading coefficients are complex Gaussian i.i.d. random variables, and in particular 
$H_e$ has covariance matrix $\Sigma_{e}=\sigma_{H_{e}}^{2}{\bf I}_{n_{e}}$.
As for the Gaussian case (as described in Section \ref{sec:gauss}), we assume that 
Alice transmits a lattice code, via coset encoding, and that the two receivers are 
performing coset decoding of the lattice, thus $n_b,n_e \geq n_t$. Indeed, if the number 
of antennas at the receiver is smaller than that of the transmitter, the lattice structure is lost at the receiver. This case will not be treated. \textcolor{black}{That $n_e \geq n_t$ might be assumed without loss of generality, since in this case Eve is in a more advantageous situation than if she had less antennas.} Finally, we denote by $\gamma_e=\sigma_{H_e}^2/\sigma_e^2$ Eve's SNR. We do not make assumption on knowing Eve's channel or on Eve's SNR, since we will compute bounds which are general, though their tightness will depend on Eve's SNR.

In order to focus on the lattice structure of the transmitted signal,
we vectorize the received signal (\ref{eq:MIMO}) and obtain
\begin{eqnarray}
\mathsf{vec}\left(Y\right) & = & \vect(H_{b}X)+\mathsf{vec}\left(V_{b}\right)\nonumber \\
 & = & \!\!\!\left(\!\!\!\begin{array}{ccc}
H_{b}\\
 & \ddots\\
 &  & H_{b}
\end{array}\!\!\!\right)\vect(X)+\vect(V_{b})\label{eq:vecy}\\
\mathsf{vec}\left(Z\right) & = & \mathsf{vec}\left(H_{e}X\right)+\mathsf{vec}\left(V_{e}\right)\nonumber \\
 & = & \!\!\!\left(\!\!\!\begin{array}{ccc}
H_{e}\\
 & \ddots\\
 &  & H_{e}
\end{array}\!\!\!\right)\vect(X)+\vect(V_{e}).\label{eq:vecz}
\end{eqnarray}
We now interpret the $n_{t}\times T$ codeword $X$ as coming from
a lattice. This is typically the case if $X$ is a space-time code coming from a division algebra \cite{Sethu}, or more 
generally if $X$ is a linear dispersion code as introduced in \cite{LDLC} where
$Tn_t$ symbols QAM are linearly encoded via a family of $Tn_t$ dispersion matrices. We write
\[
\vect(X)=M_{b}\uv
\]
 where $\uv\in\ZZ[i]^{Tn_t}$ and $M_{b}$ denotes the $Tn_{t}\times Tn_{t}$
generator matrix of the $\ZZ[i]-$lattice $\Lambda_{b}$ intended to Bob. Thus,
in what follows, by a lattice point $\xv\in\Lambda_{b}$, we mean
that
\[
\xv=\vect(X)=M_{b}\uv,
\]
 and similarly for a lattice point $\xv\in\Lambda_{e}$, we have
\[
\xv=\vect(X)=M_{e}\uv.
\]
 By setting
\begin{eqnarray*}
M_{b,H_{b}} & = & \diag(H_{b},\ldots,H_{b})M_{b},\\
M_{b,H_{e}} & = & \diag(H_{e},\ldots,H_{e})M_{b}
\end{eqnarray*}
 we can rewrite (\ref{eq:vecy}) and (\ref{eq:vecz}) as
\begin{equation}\label{eq:vecyz}
\begin{array}{lcl}
\vect(Y) & = & M_{b,H_{b}}\uv+\vect(V_{b}) \\
\vect(Z) & = & M_{b,H_{e}}\uv+\vect(V_{e}),
\end{array}
\end{equation}
where $M_{b,H_{b}}$, resp. $M_{b,H_{e}}$ can be interpreted as the lattice generators of the lattices $\Lambda_{b,H_{b}}$, resp.
$\Lambda_{b,H_{e}}$, representing the transmitted lattice seen through the respective receivers' channel, with by definition volume\footnote{Note that if 
$\Lambda$ has generator matrix $M$, we define its 
volume to be $|\det(MM^*)|^{1/2}$ if $\Lambda$ is real and $\det(MM^*)$ if $\Lambda$ is complex.}
\begin{equation}\label{eq:vol}
\begin{array}{lcl}
\vol(\Lambda_{b,H_{b}}) & = & |\det(M_{b,H_{b}}M_{b,H_{b}}^{*})|=|\det(H_{b}H_{b}^{*})|^{T}\vol(\Lambda_{b})\\
\vol(\Lambda_{b,H_{e}}) & = & |\det(M_{b,H_{e}}M_{b,H_{e}}^{*})|=|\det(H_{e}H_{e}^{*})|^{T}\vol(\Lambda_{b}).
\end{array}
\end{equation}

Similarly, the lattices $\Lambda_{e,H_{b}}$, resp. $\Lambda_{e,H_{e}}$
describe the lattices intended to Eve, seen through Bob's, resp. Eve's
channel, with respective generator matrix $M_{e,H_b}=\diag(H_b,\ldots,H_b)M_e$ and $M_{e,H_e}=\diag(H_e,\ldots,H_e)M_e$.

Note that for $\rv\in\Lambda_{e,H_{e}}$, we have
\begin{equation}\label{eq:r}
||\rv||^{2}=||\diag(H_{e},\ldots,H_e)M_{e}\uv||^{2}=||\diag(H_{e},\ldots,H_e)\xv||^{2}=||H_{e}X||_{F}^{2}
\end{equation}
where $||H_eX||_F^2=\Tr(H_eXX^*H_e^*)$ is the Frobenius norm, and $\xv=\vect(X)\in\Lambda_{e}$.

For a given realization of the channel matrices $H_e$ and $H_b$, the channel (\ref{eq:vecyz})
can be seen as the Gaussian wiretap channel
\begin{equation}
\begin{array}{ccl}
\yv & = & M_{b,H_{b}}\uv+\vv_{b}\\
\zv & = & M_{b,H_{e}}\uv+\vv_{e},
\end{array}\label{eq:chanfad}
\end{equation}
where $\yv=\vect(Y)$, $\zv=\vect(Z)$, $\vv_b=\vect(V_b)$, $\vv_e=\vect(V_e)$. We now focus on Eve's channel, since 
we know from \cite{Tarokh} how to design a good linear dispersion space-time code, \textcolor{black}{and the lattice $\Lambda_b$ is chosen so as to correspond to this space-time code}. 
We know from (\ref{eq:Pce}) that Eve's probability of correctly decoding is
\begin{eqnarray}
P_{c,e,H_{e}} & \leq & \frac{\vol(\Lambda_{b,H_{e}})}{(2\pi\sigma_{e}^{2})^{n_{t}T}}\sum_{\rv\in\Lambda_{e,H_{e}}}e^{-\Vert\rv\Vert^{2}/2\sigma_{e}^{2}}\label{eq:pce-first}\\
 & = & \frac{\vol(\Lambda_{b})}{(2\pi\sigma_{e}^{2})^{n_{t}T}}\det(H_{e}H_{e}^{*})^{T}\sum_{\xv\in\Lambda_{e}}e^{-||H_{e}X||_{F}^{2}/2\sigma_{e}^{2}}\label{eq:pce-init}
\end{eqnarray}
where last equality follows from (\ref{eq:vol}) and (\ref{eq:r}), with $\xv=\vect(X)\in\Lambda_{e}$. 
Note that as mentioned at the end of Section \ref{sec:gauss}, the exponent of $2\pi\sigma_e^2$ depends on the dimension of the transmitted lattice, which is here $n_tT$. 

Using Equation (\ref{eq:pce-init}), we derive Eve's average probability of correct decision:
\begin{eqnarray}
\bar{P}_{c,e} &=& \mathbb{E}_{H_{e}}\left[P_{c,e,H_{e}}\right]\nonumber \\
 & \leq & \frac{\vol(\Lambda_{b})}{(2\pi\sigma_{e}^{2})^{Tn_t}}\sum_{\xv\in\Lambda_{e}}\int_{\CC^{n_{e}\times n_{t}}}\det(H_{e}H_{e}^{*})^{T}
e^{-||H_{e}X||_{F}^{2}/2\sigma_{e}^{2}}\frac{e^{-\frac{1}{2}\Tr(H_{e}^{*}\Sigma_{e}^{-1}H_{e})}}{(2\pi)^{n_{e}n_{t}}\det(\Sigma_{e})^{n_{t}}}dH_{e}\nonumber\\
 & = & \frac{\vol(\Lambda_{b})}{(2\pi\sigma_{e}^{2})^{Tn_t}(2\pi)^{n_{e}n_{t}}\det(\Sigma_{e})^{n_{t}}}\nonumber \\
&&\sum_{\xv\in\Lambda_{e}}
\int_{\CC^{n_{e}\times n_{t}}}\det(H_{e}H_{e}^{*}){}^{T}e^{\frac{-1}{2\sigma_{e}^{2}}\Tr(H_{e}XX^{*}H_{e}^{*})}e^{-\frac{1}{2}\Tr(H_{e}^{*}
\Sigma_{e}^{-1}H_{e})}dH_{e}\nonumber\\
 & = & \frac{\vol(\Lambda_{b})}{(2\pi\sigma_{e}^{2})^{Tn_t}(2\pi\sigma_{H_{e}}^2)^{n_{e}n_{t}}}\nonumber \\
&&\sum_{\xv\in\Lambda_{e}}\int_{\CC^{n_{e}\times n_{t}}}\det(H_{e}H_{e}^{*})^{T}e^{-\Tr\left(H_{e}^{*}H_{e}\left[\frac{1}{2\sigma_{H_{e}}^{2}}
{\bf I}_{n_{t}}+\frac{1}{2\sigma_{e}^{2}}XX^{*}\right]\right)}dH_{e}.\label{eq:pce-devel}
\end{eqnarray}
By setting $W=H_{e}^{*}H_{e}$, we note that the above integral can be rewritten as
 \begin{equation}\label{eq:Pce-1}
 \int_{W\in\mathcal{D}_{W}}\left\{ \int_{H_{e}^{*}H_{e}=W}\det(H_{e}H_{e}^{*}){}^{T}e^{-\Tr\left(H_{e}^{*}H_{e}\left[\frac{1}{2\sigma_{H_{e}}^{2}}{\bf I}_{n_{t}}
+\frac{1}{2\sigma_{e}^{2}}XX^{*}\right]\right)}dH_{e}\right\} dW,
\end{equation}
where $\mathcal{D}_{W}$ is the set of all $n_t\times n_t$ positive definite Hermitian matrices\footnote{Note that $W$ is definite with probability 
one since $H_e$ has full rank with probability one.}.
We have, since we assumed $n_t \leq n_e$, from Theorem 2.1 in \cite{nagar-gupta} \textcolor{black}{(see also \cite{srivastava})}, that
\[
\begin{split}
\int_{H_{e}^{*}H_{e}=W}\!\!\!\!\!\det(H_{e}H_{e}^{*})^{T}e^{-\Tr\left(H_{e}^{*}H_{e}\left[\frac{1}{2\sigma_{H_{e}}^{2}}{\bf I}_{n_{t}}
+\frac{1}{2\sigma_{e}^{2}}XX^{*}\right]\right)}dH_{e}\\
=\frac{\pi^{n_{e}n_{t}}}{\Gamma_{n_{t}}(n_{e})}\left(\det W\right)^{n_{e}-n_{t}+T}e^{-\Tr\left(W\left[\frac{1}{2\sigma_{H_{e}}^{2}}{\bf I}_{n_{t}}
+\frac{1}{2\sigma_{e}^{2}}XX^{*}\right]\right)}
\end{split}
\]
where $\Gamma_{n_{t}}\left(n_{e}\right)$ is the multivariate gamma
function which can be developed as
\[
\Gamma_{n_{t}}\left(n_{e}\right)=\pi^{n_{t}\left(n_{t}-1\right)/2}\prod_{k=1}^{n_{t}}\Gamma\left(n_{e}-k+1\right).
\]
Now, Equation (\ref{eq:pce-devel}) becomes
\begin{eqnarray}
 \bar{P}_{c,e}&\leq&\frac{\vol(\Lambda_{b})\pi^{n_{e}n_{t}}}{\Gamma_{n_{t}}(n_{e})(2\pi\sigma_{e}^{2})^{n_{t}T}(2\pi\sigma_{H_{e}}^2)^{n_{e}n_{t}}}
 \sum_{\xv\in\Lambda_{e}}\!\!\int_{W\in\mathcal{D}_{W}} \!\!\!\!\!\!\!\!\!\!\det(W)^{n_{e}-n_{t}+T}
e^{-\Tr\left(W\left[\frac{1}{2\sigma_{H_{e}}^{2}}{\bf I}_{n_{t}}+\frac{1}{2\sigma_{e}^{2}}XX^{*}\right]\right)} dW \label{eq:Pce-2}\nonumber\\
&=&\frac{\vol(\Lambda_{b})\pi^{n_{e}n_{t}}\Gamma_{n_{t}}(n_{e}+T)}{\Gamma_{n_{t}}(n_{e})(2\pi\sigma_{e}^{2})^{n_{t}T}
(2\pi\sigma_{H_{e}}^2)^{n_{e}n_{t}}}\sum_{\xv\in\Lambda_{e}}\det\left(\frac{1}{2\sigma_{H_{e}}^{2}}{\bf I}_{n_{t}}+\frac{1}
{2\sigma_{e}^{2}}XX^{*}\right)^{-n_{e}-T}\label{eq:pce-3}
\end{eqnarray}
where the last equality comes from \cite{goodman}
\[
\intop_{\mathcal{D}_{W}}\left(\det W\right)^{k}\exp\left\{ -\Tr\left(\Sigma^{-1}W\right)\right\} dW=\pi^{\frac{1}{2}p(p-1)}\Gamma(p+k)\cdots\Gamma(1+k)\left(\det\Sigma\right)^{p+k}
\]
where $\mathcal{D}_{W}$ is here the set of all $p\times p$ positive definite
Hermitian matrices.

We finally obtain that an upper bound on the average probability of correct decoding for Eve is
\begin{equation}\label{eq:pce-final}
 \boxed{\bar{P}_{c,e}\leq C_{\mathrm{MIMO}}\gamma_{e}^{Tn_t}\sum_{\xv\in\Lambda_{e}}\det\left({\bf I}_{n_{t}}+
\gamma_{e}XX^{*}\right)^{-n_{e}-T}}
\end{equation}
where we set $\gamma_{e}=\frac{\sigma_{H_{e}}^{2}}{\sigma_{e}^{2}}$ for Eve's SNR,
and
\[
C_{\mathrm{MIMO}}=\frac{\vol(\Lambda_{b})\Gamma_{n_{t}}(n_{e}+T)}{\pi^{n_{t}T}\Gamma_{n_{t}}(n_{e})}.
\]

In order to design a good lattice code for the MIMO wiretap channel, we try to derive a code design criterion from Equation (\ref{eq:pce-final}):
\[
\bar{P}_{c,e}\leq C_{\mathrm{MIMO}}\gamma_{e}^{Tn_t}
\left[1+
\sum_{\xv\in\Lambda_{e}\setminus\{0\}}\det\left({\bf I}_{n_{t}}+\gamma_{e}XX^{*}\right)^{-n_{e}-T}\right]. 
\]
We can suppose that the space-time code used to transmit data to Bob is designed according to the so-called ``rank criterion'' of \cite{Tarokh}.
This means that, if $X\neq 0$ and $T\geq n_t$ then, $\mathrm{rank}(X)=n_t$. 
If we assume now that
Eve\textquoteright{}s SNR $\gamma_{e}$ is high compared to the minimum
distance of $\Lambda_{e}$, or actually design $\Lambda_e$ that way assuming Alice knows Eve's channel, we get
\begin{equation}\label{eq:pce_next}
\bar{P}_{c,e}\leq C_{\mathrm{MIMO}}
\left[\gamma_{e}^{Tn_t}+\frac{1}{\gamma_{e}^{n_en_t}}\sum_{\xv\in\Lambda_{e}\setminus\{0\}}\det\left(XX^{*}\right)^{-n_{e}-T}\right].
\end{equation}
We thus conclude that to minimize Eve's average probability of correct decoding, the design criterion is now
\begin{equation}\label{eq:criterMIMO}
\boxed{
\min_{\Lambda_e}\sum_{\xv\in\Lambda_e\setminus\{0\}}\frac{1}{\det(XX^*)^{n_e+T}}
}.
\end{equation}

\textcolor{black}{
\begin{remark}
We discuss the meaning of the bound in (\ref{eq:pce_next}). The higher $\gamma_e$, the higher should be Eve's probability of correct decoding. 
The expression in (\ref{eq:pce_next}) is decreasing as a function of $\gamma_e$ around the origin, a regime which we do not consider 
(as we just derived the expression assuming $\gamma_e$ big enough), and is then indeed increasing elsewhere as expected. 
The minimum value of this upper bound (computed by taking its derivative) is achieved for 
\[
 \gamma_{e,\min}=\left(\frac{n_e}{T}\sum_{\xv\in\Lambda_{e}\setminus\{0\}}\det\left(XX^{*}\right)^{-n_{e}-T}\right)^
{\frac{1}{n_t\left( n_e+T\right)}}. 
\]
\end{remark}}
\begin{remark}
It is important to notice that the upper bound was computed using an infinite lattice $\Lambda_e$. In some rare cases, as for an example in the case of the Alamouti code discussed later on, the bound happens to be finite even though the lattice is not. In general, it is not, in which case the bound refers not to the infinite lattice $\Lambda_e$, but instead a finite subset carved from $\Lambda_e$ via a shaping region. The same holds for the bounds derived below for block and fast fading channels. 
\end{remark}

\section{Block and Fast Fading Channels}
\label{sec:bffading}

\textcolor{black}{As a corollary of the analysis done for the MIMO case, we consider the particular fading channels where $H_{b},H_{e}$ are diagonal matrices.} In this case, setting $n_{t}=n_{b}=n_{e}=n$, the channel (\ref{eq:MIMO}) can be rewritten as 
\begin{equation}
\begin{array}{ccl}
Y & = & \diag(\hv_{b})X+V_{b}\\
Z & = & \diag(\hv_{e})X+V_{e},
\end{array}\label{eq:blockfading}
\end{equation}
which corresponds to a block fading channel with $n$ transmit antennas emitting one after the other, coherence time $T$ and
\begin{equation}
\diag(\hv_{b})=\left(\begin{array}{ccc}
h_{b,1}\\
 & \ddots\\
 &  & h_{b,n}
\end{array}\right),~\diag(\hv_{e})=\left(\begin{array}{ccc}
h_{e,1}\\
 & \ddots\\
 &  & h_{e,n}
\end{array}\right).\label{eq:fadingmatrix}
\end{equation}

However, we cannot use the final result for MIMO channels immediately, 
since the integral over all positive definite Hermitian matrices does not hold anymore. Moreover, the general 
expression of (\ref{eq:Pce-1}) does not hold either since it assumes that $H_e$ (here $\diag(\hv_e)$) is i.i.d distributed. 
We thus start from the generic equation (\ref{eq:pce-init}), which gives, using a polar coordinates change, and the change 
of variables $u_{e,i}=\rho_{e,i}^2$
\begin{eqnarray*}
\bar{P}_{c,e} &\leq &\frac{\vol(\Lambda_{b})}{(2\pi\sigma_{e}^{2})^{nT}(2\pi\sigma_{\hv_{e}}^2)^{n}}
\sum_{\xv\in\Lambda_{e}}\prod_{i=1}^n\int_{\CC}\left|h_{e,i}\right|^{2T}
e^{-\left|h_{e,i}\right|^2\left[\frac{1}{2\sigma_{\hv_{e}}^{2}}+\frac{1}{2\sigma_{e}^{2}}||\xv_i||^2\right]} dh_{e,i} \\
& =&\frac{(2\pi)^n \vol(\Lambda_{b})}{(2\pi\sigma_{e}^{2})^{nT}(2\pi\sigma_{\hv_{e}}^2)^{n}}
\sum_{\xv\in\Lambda_{e}}\prod_{i=1}^n\int_{0}^{\infty}\rho_{e,i}^{2T+1}
e^{-\rho_{e,i}^2\left[\frac{1}{2\sigma_{\hv_{e}}^{2}}+\frac{1}{2\sigma_{e}^{2}}||\xv_i||^2\right]} d\rho_{e,i} \\
& =&\frac{\vol(\Lambda_{b})}{(2\pi\sigma_{e}^{2})^{nT}(2\sigma_{\hv_{e}}^2)^{n}}
\sum_{\xv\in\Lambda_{e}}\prod_{i=1}^n\int_{0}^{\infty}u_{e,i}^{T}
e^{-u_{e,i}\left[\frac{1}{2\sigma_{\hv_{e}}^{2}}+\frac{1}{2\sigma_{e}^{2}}||\xv_i||^2\right]} du_{e,i} \\
& =&\frac{\Gamma(1+T)^n\vol(\Lambda_{b})}{(2\pi\sigma_{e}^{2})^{nT}(2\sigma_{\hv_{e}}^2)^{n}}
\sum_{\xv\in\Lambda_{e}}\prod_{i=1}^n\left[\frac{1}{2\sigma_{\hv_{e}}^{2}}+\frac{1}{2\sigma_{e}^{2}}||\xv_i||^2\right]^{-1-T}.
\end{eqnarray*}
We finally obtain an upper bound of the average probability of correct decision for Eve for the wiretap block fading channel, 
given by
\begin{equation}\label{eq:Pce-BF}
\boxed{\bar{P}_{c,e}\leq C_{\mathrm{BF}}\gamma_e^{nT}
\sum_{\xv\in\Lambda_{e}}\prod_{i=1}^n\left[1+\gamma_e||\xv_i||^2\right]^{-1-T}}
\end{equation}
where 
\[
 C_{\mathrm{BF}}=\frac{(T!)^n\vol(\Lambda_{b})}{\pi^{nT}}
\]
and similarly to the MIMO case, $\gamma_e=\frac{\sigma_{\hv_e}^2}{\sigma_e^2}$.

In order to design a good lattice code for the block fading wiretap channel, 
we now try to derive a code design criterion from (\ref{eq:Pce-BF}):
\[
\bar{P}_{c,e}\leq C_{\mathrm{BF}}\gamma_{e}^{Tn}
\left[1+
\sum_{\xv\in\Lambda_{e}\setminus\{0\}}\prod_{i=1}^n\left[1+\gamma_e||\xv_i||^2\right]^{-1-T}\right]. 
\]
We can suppose that the code used to transmit data to Bob is designed according to the minimum product distance criterion.
This means that, if $\xv\neq 0$, then, $\xv_i\neq 0$ for any $i$. 
If we assume this time that
Eve\textquoteright{}s SNR $\gamma_{e}$ is high compared to the minimum
distance of $\Lambda_{e}$, or actually design $\Lambda_e$ that way assuming Alice knows Eve's channel, we get
\[
\bar{P}_{c,e}\leq C_{\mathrm{BF}}
\left[\gamma_{e}^{Tn}+\frac{1}{\gamma_{e}^{n}}\sum_{\xv\in\Lambda_{e}\setminus\{0\}}
\prod_{i=1}^n\left(||\xv_i||^2\right)^{-1-T}\right].
\]
This expression is decreasing as a function of $\gamma_e$ around the origin, a regime which we do not consider 
(as we again just derived the expression assuming $\gamma_e$ big enough), and is then indeed increasing as expected. 
The minimum value of this upper bound is achieved for 
\[
 \gamma_{e,\min}=\left(\frac{\sum_{\xv\in\Lambda_{e}\setminus\{0\}}\prod_{i=1}^n\left(||\xv_i||^2\right)^{-1-T}}{T}\right)^
{\frac{1}{n\left( 1+T\right)}}. 
\]

We thus conclude that to minimize Eve's average probability of correct decoding, the design criterion is now
\[
\boxed{
\min_{\Lambda_e}\sum_{\xv\in\Lambda_e\setminus\{0\}}\frac{1}{\left(\prod_{i=1}^n||\xv_i||^2\right)^{1+T}}
}.
\]

When furthermore $T=1$ (and $X$ is thus a $n\times1$ vector $\xv$) in (\ref{eq:blockfading}), we get a fast fading channel:
\begin{equation*}
\begin{array}{ccl}
\yv & = & \diag(\hv_{b})\xv+\vv_{b}\\
\zv & = & \diag(\hv_{e})\xv+\vv_{e},
\end{array}
\end{equation*}
where all vectors are $n$-dimensional complex vectors corresponding to $n$ usages of the channel, and
\begin{equation*}
\diag(\hv_{b})=\left(\begin{array}{ccc}
h_{b,1}\\
 & \ddots\\
 &  & h_{b,n}
\end{array}\right),~\diag(\hv_{e})=\left(\begin{array}{ccc}
h_{e,1}\\
 & \ddots\\
 &  & h_{e,n}
\end{array}\right)
\end{equation*}
as before (see (\ref{eq:fadingmatrix})). 
We can thus immediately apply the result (\ref{eq:Pce-BF}) to deduce that 
\begin{equation}\label{eq:Pce-FF}
\boxed{\bar{P}_{c,e}\leq C_{\mathrm{FF}}\gamma_e^{n}
\sum_{\xv\in\Lambda_{e}}\prod_{i=1}^n\left[1+\gamma_e|x_i|^2\right]^{-2}}
\end{equation}
where 
\[
 C_{\mathrm{FF}}=\frac{\vol(\Lambda_{b})}{\pi^{n}}
\]
and still again, $\gamma_e=\frac{\sigma_{\hv_e}^2}{\sigma_e^2}$. 
The design criterion follows accordingly
\[
\boxed{
\min_{\Lambda_e}\sum_{\xv\in\Lambda_e\setminus\{0\}}\frac{1}{\left(\prod_{i=1}^n|x_i|^2\right)^{2}}
}.
\]
We thus recover the expressions presented in \cite{icc}, though here in the complex case, which explains 
the difference in the exponent \footnote{Please note an erratum in \cite{icc}, since the sum derived there is 
over all lattice points, while of course, the zero vector should be removed from the sum.}.



\section{A MIMO example: The Alamouti Code}
\label{sec:alamouti}

In this section, we illustrate the code design criterion derived above using the Alamouti code \cite{Alamouti} 
with QAM constellation, $n_{t}=2$, $n_{e}\geq2$ and $T=2$. \textcolor{black}{Note that the Alamouti code does not form a $\mathbb{Z}[i]$-lattice, but a $\ZZ$-lattice. We choose the Alamouti code nevertheless since this is the best understood and the simplest MIMO code available in the literature. It is not difficult to check that our analysis, and thus the resulting code design, holds for real lattices as well.}
An Alamouti codeword is then of the form
\[
X=\left[\begin{array}{cc}
x_{1} & x_{2}\\
-x_2^* & x_{1}^*
\end{array}\right],~x_1,x_2\in\ZZ[i],
\]
so that
\[
\det\left(XX^{*}\right)=\left(\left|x_{1}\right|^{2}+\left|x_{2}\right|^{2}\right)^{2}=\left\Vert \mathbf{x}\right\Vert ^{4},
\]
where 
\[
\mathbf{x}=\left[\begin{array}{c}
x_{1}\\
x_{2}
\end{array}\right]\in\ZZ[i]^2=\Lambda_e.
\]
The design criterion (\ref{eq:criterMIMO}) requires to study
\begin{equation}\label{eq:expr-psi}
\sum_{\mathbf{x}\in\Lambda_{e}\setminus\left\{ \mathbf{0}\right\} }\det\left(XX^{*}\right)^{-n_{e}-T}
=
\sum_{\mathbf{x}\in\Lambda_{e}\setminus\{\mathbf{0}\}}\frac{1}{\left\Vert \xv\right\Vert ^{2\left(2\left(n_{e}+T\right)\right)}}=\zeta_{\Lambda_{e}}\left(2\left(n_{e}+2\right)\right),
\end{equation}
where we recognize the Epstein zeta function of a scaled lattice $\mu\Lambda$ ($\mu>0$), defined by
\begin{equation}\label{eq:zeta-scaled}
\zeta_{\mu\Lambda}(s)=\sum_{\mathbf{x}\in\Lambda\setminus\left\{ \mathbf{0}\right\} }\frac{1}{\mu^{2s}}\frac{1}{\left\Vert \mathbf{x}\right\Vert ^{2s}}=\frac{1}{\mu^{2s}}\zeta_{\Lambda}(s).
\end{equation}

Since $\xv\in\ZZ[i]^2\simeq \ZZ^4$, we will consider as possible lattices $\Lambda_e$ either $\ZZ^4$ itself, with 
Epstein zeta function (see Proposition \ref{prop:zeta-z4} in Appendix)
\begin{equation}\label{eq:zeta-z4}
\zeta_{\mathbb{Z}^{4}}(s) =  8\left(1-4^{1-s}\right)\zeta(s)\zeta(s-1)
\end{equation}
or $D_4$, in which case the vector $\xv$ above is coded, and belongs to $D_4$\footnote{The complex construction $D_4=(1+i)\ZZ[i]^2+(2,1,2)$ may be used, for instance, 
where $(2,1,2)$ is the repetition code of length $2$.} 
instead of $\ZZ^4$, which in turn involved 
the Epstein zeta function of $D_4$ (see Proposition \ref{prop:zeta-d4} also in Appendix)
\begin{equation}\label{eq:zeta-d4}
\zeta_{D_{4}}(s)= 3\cdot4^{2-s}\left(2^{s-1}-1\right)\zeta(s)\zeta(s-1).
\end{equation}
In both cases, $\zeta(s)=\sum_{n>0}\frac{1}{n^s}$ is the Riemann zeta function. 

In order to compare the Epstein zeta function of the two lattices $\mathbb{Z}^{4}$ and $D_{4}$,
we rescale $D_{4}$ so that its fundamental volume is equal to the
fundamental volume of $\mathbb{Z}^{4}$, that is $1$. Since $\mathrm{vol}\left(D_{4}\right)=2$,
the scaling factor is $\mu=\frac{1}{\sqrt[4]{2}}$. Combining (\ref{eq:zeta-d4}) and (\ref{eq:zeta-scaled}),
we obtain 
\begin{eqnarray}
\zeta_{\frac{1}{\sqrt[4]{2}}D_{4}}(s) & = & \left(\sqrt[4]{2}\right)^{2s}3\cdot4^{2-s}\left(2^{s-1}-1\right)\zeta(s)\zeta(s-1)\nonumber \\
 & = & 3\cdot2^{4-3\frac{s}{2}}\cdot\left(2^{s-1}-1\right)\zeta(s)\zeta(s-1)\label{eq:zeta-d4-normal}
\end{eqnarray}
where $s=2n_{e}+4$, which we have to compare with 
\[
\zeta_{\mathbb{Z}^{4}}(s)=8\left(1-4^{1-s}\right)\zeta(s)\zeta(s-1).
\]
We eventually define the gain $\varsigma_{D_{4}}$ obtained by using $D_{4}$ instead of
$\mathbb{Z}^{4}$ (the uncoded case) as 
\begin{eqnarray*}
\varsigma_{D_{4}}
&=& \frac{\zeta_{\mathbb{Z}^{4}}(s)}{\zeta_{\mu D_{4}}(s)}|_{s=2n_e+4}\\
&=&  \frac{-2^3 4^{1-s}(1-4^{s-1})}{3\cdot2^{4-3\frac{s}{2}}\cdot\left(2^{s-1}-1\right)}\\
& = & \left.\frac{1}{3\cdot 2^{\frac{s}{2}-1}}\left(2^{s-1}+1\right)\right|_{s=2n_{e}+4}\\ 
& = & \frac{2^{2n_{e}+3}+1}{3\cdot2^{n_{e}+1}}\cong\frac{4}{3}2^{n_{e}}.
\end{eqnarray*}

\begin{figure}[ht]
\noindent \begin{centering}
\includegraphics[width=12cm]{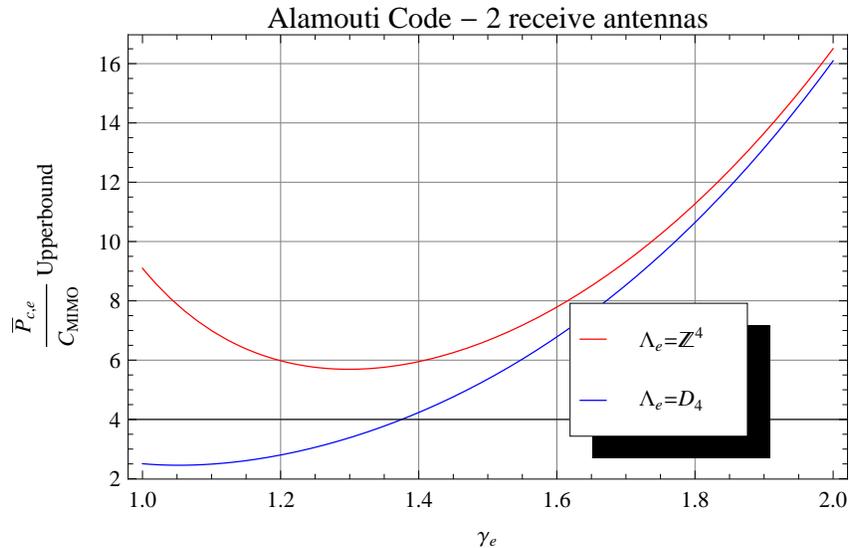}
\par\end{centering}
\caption{\label{fig:-upperbound:-Alamouti} An upper bound on $\tfrac{\bar{P}_{c,e}}{C_{\mathrm{MIMO}}}$: the Alamouti
code with $n_{e}=2$.}
\end{figure}

We illustrate the obtained results on Figure \ref{fig:-upperbound:-Alamouti} by plotting the upper bound (\ref{eq:pce_next}) on Eve's probability $\bar{P}_{c,e}$ of correct decision, divided by the constant $C_{\rm{MIMO}}$, when the Alamouti code is used with
as coarse lattice $\Lambda_e$ either $\mathbb{Z}^{4}$ or $D_{4}$. 

Notice that when $\gamma_{e}$ is small, this upper bound becomes of course very loose as
it is a decreasing function in $\gamma_e$, while we expect on the contrary $\bar{P}_{c,e}\left(\gamma_{e}\right)$ to be an increasing function. 
This motivates the following discussion the tightness of the upper bound (\ref{eq:pce_next}). 

We go back to the tighter upper bound (\ref{eq:pce-final}) on $\bar{P}_{c,e}$: 
\begin{eqnarray}
\bar{P}_{c,e} & \le & C_{\mathrm{MIMO}}\gamma_{e}^{Tn_{t}}\sum_{\xv\in\Lambda_{e}}\det\left({\bf I}_{n_{t}}+\gamma_{e}XX^{*}\right)^{-n_{e}-T}\nonumber \\
 & = & C_{\mathrm{MIMO}}\underbrace{\gamma_{e}^{-n_{e}n_{t}}\sum_{\xv\in\Lambda_{e}}\det\left(\frac{1}{\gamma_{e}}{\bf I}_{n_{t}}+XX^{*}\right)^{-n_{e}-T}}_{\varphi_{\Lambda_{e}}\left(\gamma_{e}\right)}.\label{eq:pce_alamouti}
\end{eqnarray}
When $X$ is a codeword from the Alamouti code, with $T=n_t=2$, then 
\[
\det\left(\frac{1}{\gamma_{e}}{\bf I}_{n_{t}}+XX^{*}\right)=\left(\frac{1}{\gamma_{e}}+\left\Vert \mathbf{x}\right\Vert ^{2}\right)^{2},
\]
so that
\begin{eqnarray}
\varphi_{\Lambda_{e}}\left(\gamma_{e}\right) & \triangleq & \gamma_{e}^{-n_{e}n_{t}}\sum_{\xv\in\Lambda_{e}}\det\left(\frac{1}{\gamma_{e}}{\bf I}_{n_{t}}+XX^{*}\right)^{-n_{e}-T}\\
 & = & \gamma_{e}^{-2n_{e}}\sum_{\xv\in\Lambda_{e}}\frac{1}{\left(\frac{1}{\gamma_{e}}+\left\Vert \mathbf{x}\right\Vert ^{2}\right)^{2\left(n_{e}+2\right)}}\\
 & = & \gamma_{e}^{4}+\gamma_{e}^{-2n_{e}}\sum_{\xv\in\Lambda_{e}\setminus\left\{ \boldsymbol{0}\right\} }\frac{1}{\left(\frac{1}{\gamma_{e}}+\left\Vert \mathbf{x}\right\Vert ^{2}\right)^{2\left(n_{e}+2\right)}}.\label{eq:sum-det-hurwitz}
\end{eqnarray}

We are thus interested in the calculation of 
\[
\zeta_{\Lambda_e}\left(s,a\right)\triangleq\sum_{\mathbf{x}\in\Lambda_e}\frac{1}{\left(a+\left\Vert \mathbf{x}\right\Vert ^{2}\right)^{s}}=\frac{1}{a^{s}}+\sum_{\mathbf{x}\in\Lambda_e\setminus\left\{ \boldsymbol{0}\right\} }\frac{1}{\left(a+\left\Vert \mathbf{x}\right\Vert ^{2}\right)^{s}},~a=\frac{1}{\gamma_e},~s=2(n_e+2),
\]
which will be done via the Mellin transform
\[
\mathcal{M}(f)(s) = \int_{0}^{+\infty} f(t) t^{s-1} dt,
\] 
thanks to which we obtain the following:
\begin{lem}
If $a<||\xv||^2$, we have that 
\[
\frac{1}{\left(a+\left\Vert \mathbf{x}\right\Vert ^{2}\right)^{s}}\\
=
\sum_{k=0}^{+\infty}
\frac{\left(-1\right)^{k}a^{k}}{k!}\frac{\Gamma(s+k)}{\Gamma(s)}
\left(\frac{1}{\left\Vert \mathbf{x}\right\Vert ^{2}}\right)^{s+k}.
\]
\end{lem}
\begin{IEEEproof}
The Mellin transform of $e^{-(a+\left\Vert \mathbf{x}\right\Vert ^{2})t}$, for some positive $a\in\RR$, is 
\begin{eqnarray}
\mathcal{M}\left(e^{-(a+\left\Vert \mathbf{x}\right\Vert ^{2})t}\right)(s)  
  &=&  \int_{0}^{+\infty}e^{-(a+\left\Vert \mathbf{x}\right\Vert ^{2})t}t^{s}\frac{dt}{t} \nonumber\\
  &=&  \frac{1}{(a+\left\Vert \mathbf{x}\right\Vert ^{2})^{s}}\int_{0}^{+\infty}e^{-u}u^{s}\frac{du}{u}\nonumber \\
  &=&\Gamma(s)\frac{1}{\left(a+\left\Vert \mathbf{x}\right\Vert ^{2}\right)^{s}},\label{eq:mellin-shift-exp}
\end{eqnarray}
which we can alternatively write as
\begin{eqnarray}
\mathcal{M}(e^{-\left(a+\left\Vert \mathbf{x}\right\Vert ^{2}\right)t})(s)
&=&
\mathcal{M}(e^{-at}e^{-\left\Vert \mathbf{x}\right\Vert ^{2}t})(s)\nonumber\\
&=&
\int_0^{+\infty}\sum_{k=0}^{+\infty}\frac{\left(-1\right)^{k}a^{k}}{k!}t^{k+s}e^{-\left\Vert \mathbf{x}\right\Vert ^{2}t}\frac{dt}{t}\label{eq:switch-sum} \\
&=&
\sum_{k=0}^{+\infty}
\frac{\left(-1\right)^{k}a^{k}}{k!}\Gamma(s+k)\left(\frac{1}{\left\Vert \mathbf{x}\right\Vert ^{2}}\right)^{s+k}\nonumber.
\end{eqnarray}
Now (\ref{eq:switch-sum}) involves a dangerous exchange of an integral with an infinite sum. For it to be allowed, we need to check that
\[
\left(\sum_{k=0}^n \int_{0}^{\infty} \frac{|\left(-1\right)^{k}|a^{k}}{k!}t^{k+s}e^{-\left\Vert \mathbf{x}\right\Vert ^{2}t}\frac{dt}{t} \right)_{n\in\mathbb{N}}
\]
converges for every $t$ in the integration range, which is the same as showing that
\[
\left(\sum_{k=0}^n  \frac{|\left(-1\right)^{k}|a^{k}}{k!}\Gamma(s+k)\left(\frac{1}{\left\Vert \mathbf{x}\right\Vert ^{2}}\right)^{s+k}
\right)_{n\in\mathbb{N}}
\]
converges for every $t$ in the integration range. Since we only have strictly positive terms, comparing the $n$th term with the $(n+1)$th term yields, recalling that since $s=2(n_2+2)$, $\Gamma(s+n)=(s+n-1)!$:
\[
\frac{a}{n+1}\frac{s+n}{||\xv||^2}
\]
whose limit needs to be stricly smaller than 1, that is 
\begin{equation}\label{eq:cond-conv}
\lim_{n\rightarrow\infty} \frac{a}{n+1}\frac{s+n}{||\xv||^2}=\frac{a}{||\xv||^2} < 1,
\end{equation}
showing that the above computation is valid when $a< ||\xv||^2$.
We then have, comparing (\ref{eq:mellin-shift-exp}) and  (\ref{eq:switch-sum}), that 
\[
\frac{1}{\left(a+\left\Vert \mathbf{x}\right\Vert ^{2}\right)^{s}}\\
=
\sum_{k=0}^{+\infty}
\frac{\left(-1\right)^{k}a^{k}}{k!}\frac{\Gamma(s+k)}{\Gamma(s)}
\left(\frac{1}{\left\Vert \mathbf{x}\right\Vert ^{2}}\right)^{s+k}.
\]
\end{IEEEproof}

We are now ready to prove the following result for $\ZZ^4$. The equivalent result for $D_4$ follows, and the consequences of both computations for the bound on the error probability can be found below in Corollary \ref{cor:pce}.
\begin{prop}
Suppose $0<a<q$.
For the lattice $\ZZ^4$, we have
\[
\zeta_{\ZZ^4}\left(s,a\right) =
\sum_{\mathbf{x}\in\ZZ^4}\frac{1}{\left(a+\left\Vert \mathbf{x}\right\Vert ^{2}\right)^{s}}
=\sum_{j=0}^{q-1} \frac{r_4(j)}{(a+j)^s}+
\sum_{k=0}^{+\infty}\frac{\left(-1\right)^{k}a^k}{k!}\frac{\Gamma(s+k)}
{\Gamma(s)}\left( -8\cdot 4^{1-s-k}-\sum_{j=2}^{q-1} \frac{r_4(j)}{j^{s+k}} \right),
\]
where $r_4(j)$ denotes the number of vectors of norm $j$ in $\ZZ^4$. 
In particular if $0<a<2$, we have
\[
\zeta_{\ZZ^4}\left(s,a\right) =
\sum_{\mathbf{x}\in\ZZ^4}\frac{1}{\left(a+\left\Vert \mathbf{x}\right\Vert ^{2}\right)^{s}}=
\frac{1}{a^s}+\frac{8}{(1+a)^s}-8\cdot 4^{1-s}\left(1-\frac{a}{4}\right)^s,
\]
and
\[
\zeta_{\ZZ^4}\left(s,a\right) \leq
\sum_{j=0}^{3} \frac{r_4(j)}{(a+j)^s}-8\cdot 4^{1-s}\left(1-\frac{a}{4}\right)^s-
\frac{1}{4^s}\sum_{j=2}^{3} r_4(j) \left(1-\frac{a}{4}\right)^s
\]
if $0<a<4$.
\end{prop}
\begin{IEEEproof} 
Since $\ZZ^4$ is an integer lattice with vectors of norm 1, we have $||\xv||^2 \geq 1$ if $\xv\neq 0$, that is we need  $a<1$ to use the above lemma. Alternatively, if we consider lattice points whose norm is at least $q$, we 
can use $a<q$,  which gives
\begin{eqnarray*}
\sum_{\mathbf{x}\in\Lambda_e}\frac{1}{\left(a+\left\Vert \mathbf{x}\right\Vert ^{2}\right)^{s}}
&=&
\sum_{j=0}^{q-1} \frac{r_4(j)}{(a+j)^s}+
\sum_{\mathbf{x}\in\Lambda_e, ||\xv||^2\geq q }\frac{1}{\left(a+\left\Vert \mathbf{x}\right\Vert ^{2}\right)^{s}}\\
&=&
\sum_{j=0}^{q-1} \frac{r_4(j)}{(a+j)^s}+
\sum_{\mathbf{x}\in\Lambda_e,||\xv||^2\geq q }
\sum_{k=0}^{+\infty}
\frac{\left(-1\right)^{k}a^{k}}{k!}\frac{\Gamma(s+k)}{\Gamma(s)}
\left(\frac{1}{\left\Vert \mathbf{x}\right\Vert ^{2}}\right)^{s+k}\\
&=&
\sum_{j=0}^{q-1} \frac{r_4(j)}{(a+j)^s}+
\sum_{k=0}^{+\infty}
\frac{\left(-1\right)^{k}a^{k}}{k!}\frac{\Gamma(s+k)}{\Gamma(s)}
\sum_{\mathbf{x}\in\Lambda_e,||\xv||^2\geq q}
\frac{1}{\left\Vert \mathbf{x}\right\Vert ^{2(s+k)}}\\
&=&
\sum_{j=0}^{q-1} \frac{r_4(j)}{(a+j)^s}+
\sum_{k=0}^{+\infty}\frac{\left(-1\right)^{k}a^k}{k!}\frac{\Gamma(s+k)}
{\Gamma(s)}\left(\zeta_{\Lambda_e}(s+k)- \sum_{j=1}^{q-1} \frac{r_4(j)}{j^{s+k}}\right)
\end{eqnarray*}
where $\zeta_{\Lambda_e}(s)$ is the Epstein zeta function of $\Lambda$ defined in (\ref{eq:zeta-scaled}), and $r_4(j)$ counts the number of vectors of norm $j$ in $\ZZ^4$. 
We were allowed to exchange both infinite sums since $\zeta_{\Lambda_e}(s+k)$ converges, and thus so does $\zeta_{\Lambda_e}(s+k)-\sum_{j=1}^{q-1} \frac{r_4(j)}{j^{s+k}}$, for every $k \geq 0$, and 
\begin{equation}\label{eq:conv}
\sum_{k=0}^{+\infty}\frac{|\left(-1\right)^{k}|a^k}{k!}\frac{\Gamma(s+k)}
{\Gamma(s)}\left(\zeta_{\Lambda_e}(s+k)-\sum_{j=1}^{q-1} \frac{r_4(j)}{j^{s+k}}\right)
\end{equation}
converges as well, which follows from
\[
\frac{a(s+n)\left(\zeta_{\Lambda_e}(s+n+1)-\sum_{j=1}^{q-1} \frac{r_4(j)}{j^{s+n+1}}\right)}{(n+1)\left(\zeta_{\Lambda_e}(s+n)-\sum_{j=1}^{q-1} \frac{r_4(j)}{j^{s+n}}\right)} \leq 
\frac{a(s+n)}{q(n+1)} \rightarrow \frac{a}{q} <1
\]
when $n$ grows, noting that
\[
\zeta_{\Lambda_e}(s+n+1)-\sum_{j=1}^{q-1} \frac{r_4(j)}{j^{s+n+1}}
=\sum_{\mathbf{x}\in\Lambda_e,||\xv||^2\geq q}
\frac{1}{\left\Vert \mathbf{x}\right\Vert ^{2(s+n)}||\xv||^2}\leq  \frac{1}{q}\sum_{\mathbf{x}\in\Lambda_e,||\xv||^2\geq q}
\frac{1}{\left\Vert \mathbf{x}\right\Vert ^{2(s+n)}}. 
\]

The Epstein zeta function of the lattice $\ZZ^4$ is given (see Proposition \ref{prop:zeta-z4}) by 
\[
\zeta_{\ZZ^4}(s)=8(1-4^{1-s})\zeta(s)\zeta(s-1)
\]
and since $s=2(n_e+2)\geq 8$ when $n_2\geq 2$, $\zeta(s+k)\zeta(s-1+k)$ can be approximated by the smallest value of $k$ and $s$, namely $\zeta(8)\zeta(7)$, where $\zeta(7)\simeq 1.00835$. Thus
\[
\zeta_{\ZZ^4}(s+k)= 8(1-4^{1-s-k})\zeta(s+k)\zeta(s+k-1)\simeq 8(1-4^{1-s-k}),
\] 
and, using that $r_4(1)=8$ (there are 8 vectors of norm 1, the 4 unit vectors and the same vectors with a minus sign)
\begin{eqnarray*}
\sum_{\mathbf{x}\in\ZZ^4}\frac{1}{\left(a+\left\Vert \mathbf{x}\right\Vert ^{2}\right)^{s}}
&=&
\sum_{j=0}^{q-1} \frac{r_4(j)}{(a+j)^s}+
\sum_{k=0}^{+\infty}\frac{\left(-1\right)^{k}a^k}{k!}\frac{\Gamma(s+k)}
{\Gamma(s)}\left( 8(1-4^{1-s-k})-\sum_{j=1}^{q-1} \frac{r_4(j)}{j^{s+k}} \right) \\
&=&
\sum_{j=0}^{q-1} \frac{r_4(j)}{(a+j)^s}+
\sum_{k=0}^{+\infty}\frac{\left(-1\right)^{k}a^k}{k!}\frac{\Gamma(s+k)}
{\Gamma(s)}\left( -8\cdot 4^{1-s-k}-\sum_{j=2}^{q-1} \frac{r_4(j)}{j^{s+k}} \right). 
\end{eqnarray*}

In particular if $q=2$, we get that 
\begin{eqnarray*}
\sum_{\mathbf{x}\in\ZZ^4}\frac{1}{\left(a+\left\Vert \mathbf{x}\right\Vert ^{2}\right)^{s}}
&=&
\frac{1}{a^s}+\frac{8}{(1+a)^s}-8\cdot 4^{1-s}
\sum_{k=0}^{+\infty}\frac{\left(-1\right)^{k}a^k}{4^k}\frac{\Gamma(s+k)}
{k!\Gamma(s)}\\
&=&
\frac{1}{a^s}+\frac{8}{(1+a)^s}-8\cdot 4^{1-s}
\sum_{k=0}^{+\infty}\left(\frac{-a}{4}\right)^{k} {s+k-1 \choose k} \\
&=&
\frac{1}{a^s}+\frac{8}{(1+a)^s}-8\cdot 4^{1-s}
\sum_{k=0}^{+\infty}\left(\frac{-a}{4}\right)^{k} {s-1 \choose k},
\end{eqnarray*}
recalling the definition of Gamma functions for positive integers.
In summary, recognizing the generalized binomial coefficients, we get
\[
\sum_{\mathbf{x}\in\ZZ^4}\frac{1}{\left(a+\left\Vert \mathbf{x}\right\Vert ^{2}\right)^{s}}=
\frac{1}{a^s}+\frac{8}{(1+a)^s}-8\cdot 4^{1-s}\left(1-\frac{a}{4}\right)^s.
\]

If instead $q=4$, we note first (this first inequality holds for any $q$ but not what will follow) that 
\[
\sum_{j=2}^{q-1} \frac{r_4(j)}{j^{s+k}} \geq  \frac{1}{q^{s+k}}\sum_{j=2}^{q-1} r_4(j),
\]
so that
\begin{eqnarray*}
&&\sum_{\mathbf{x}\in\ZZ^4}\frac{1}{\left(a+\left\Vert \mathbf{x}\right\Vert ^{2}\right)^{s}}\\
&\leq&
\sum_{j=0}^{q-1} \frac{r_4(j)}{(a+j)^s}-8\cdot 4^{1-s}
\sum_{k=0}^{+\infty}\frac{\left(-1\right)^{k}a^k}{4^k}\frac{\Gamma(s+k)}
{k!\Gamma(s)}- \frac{1}{q^s}\sum_{j=2}^{q-1} r_4(j) 
\sum_{k=0}^{+\infty}\frac{\left(-1\right)^{k}a^k}{q^k}\frac{\Gamma(s+k)}
{k!\Gamma(s)}  \\
&=&
\sum_{j=0}^{q-1} \frac{r_4(j)}{(a+j)^s}-8\cdot 4^{1-s}\left(1-\frac{a}{4}\right)^s-
\frac{1}{q^s}\sum_{j=2}^{q-1} r_4(j) \left(1-\frac{a}{q}\right)^s.
\end{eqnarray*}
The condition $q=4$ ensures the convergence of the second series.
\end{IEEEproof}

\begin{prop}
Suppose $0<a<q$. For the lattice $D_4$, we have 
\[
\zeta_{D_4}\left(s,a\right) 
=\sum_{\mathbf{x}\in D^4}\frac{1}{\left(a+\left\Vert \mathbf{x}\right\Vert ^{2}\right)^{s}}
=\sum_{j=0}^{q-1} \frac{r_{D_4}(j)}{(a+j)^s}+
\sum_{k=0}^{+\infty}\frac{\left(-1\right)^{k}a^k}{k!}\frac{\Gamma(s+k)}
{\Gamma(s)}\left(-3\cdot4^{2-s-k} - \sum_{j=3}^{q-1} \frac{r_{D_4}(j)}{j^{s+k}}\right),
\]
where $r_{D_4}(j)$ denotes the number of vectors of length $j$ in $D_4$. 
In particular, if $0<a<2$, we have
\[
\zeta_{D_4}\left(s,a\right) =
\frac{1}{a^s}
-3\cdot 4^{2-s} \left(1-\frac{a}{4}\right)^s,
\]
and
\[
\zeta_{D_4}\left(s,a\right) =
\frac{1}{a^s}+\frac{24}{(a+2)^s}
-3\cdot 4^{2-s} \left(1-\frac{a}{4}\right)^s
\]
for $0<a<4$.
\end{prop}
\begin{IEEEproof}
The following computed above for $\ZZ^4$ holds similarly for $D_4$
\[
\sum_{\mathbf{x}\in D^4}\frac{1}{\left(a+\left\Vert \mathbf{x}\right\Vert ^{2}\right)^{s}}=
\sum_{j=0}^{q-1} \frac{r_{D_4}(j)}{(a+j)^s}+
\sum_{k=0}^{+\infty}\frac{\left(-1\right)^{k}a^k}{k!}\frac{\Gamma(s+k)}
{\Gamma(s)}\left(\zeta_{\Lambda_e}(s+k)- \sum_{j=1}^{q-1} \frac{r_{D_4}(j)}{j^{s+k}}\right)
\]
where we use the notation $r_{D_4}(j)$ to denote the number of vectors of length $j$ in $D_4$. 
Note that $D_4$ has no vector of norm 1, and 24 of norm 2.

From Proposition \ref{prop:zeta-d4}, the Epstein zeta function of $D_4$ is 
\[
\zeta_{D_{4}}(s+k)  =  3\cdot4^{2-s-k}\left(2^{s+k-1}-1\right)\zeta(s+k)\zeta(s+k-1),
\]
and as before for $\ZZ^4$
\[
\zeta_{D_{4}}(s+k)  \simeq 3\cdot4^{2-s-k}\left(2^{s+k-1}-1\right),
\]
so that
\begin{eqnarray*}
\sum_{\mathbf{x}\in D_4}\frac{1}{\left(a+\left\Vert \mathbf{x}\right\Vert ^{2}\right)^{s}}
&=&
\sum_{j=0}^{q-1} \frac{r_{D_4}(j)}{(a+j)^s}+
\sum_{k=0}^{+\infty}\frac{\left(-1\right)^{k}a^k}{k!}\frac{\Gamma(s+k)}
{\Gamma(s)}\left(3\cdot4^{2-s-k}\left(2^{s+k-1}-1\right) - \sum_{j=1}^{q-1} \frac{r_{D_4}(j)}{j^{s+k}}\right)\\
&=&
\sum_{j=0}^{q-1} \frac{r_{D_4}(j)}{(a+j)^s}+
\sum_{k=0}^{+\infty}\frac{\left(-1\right)^{k}a^k}{k!}\frac{\Gamma(s+k)}
{\Gamma(s)}\left(-3\cdot4^{2-s-k} - \sum_{j=3}^{q-1} \frac{r_{D_4}(j)}{j^{s+k}}\right).
\end{eqnarray*}

If $q=2$, we can simplify the above expression to get
\begin{eqnarray*}
\sum_{\mathbf{x}\in D_4}\frac{1}{\left(a+\left\Vert \mathbf{x}\right\Vert ^{2}\right)^{s}}
&=&
\frac{1}{a^s}+
\sum_{k=0}^{+\infty}\frac{\left(-1\right)^{k}a^k}{k!}\frac{\Gamma(s+k)}
{\Gamma(s)}\left(-3\cdot4^{2-s-k} \right)\\
&=&
\frac{1}{a^s}-3\cdot4^{2-s}
\sum_{k=0}^{+\infty}\frac{\left(-1\right)^{k}a^k}{4^k}\frac{\Gamma(s+k)}
{k!\Gamma(s)}\\
&=&
\frac{1}{a^s}
-3\cdot 4^{2-s} \left(1-\frac{a}{4}\right)^s,
\end{eqnarray*}
while if $q=4$, recalling that $D_4$ has no vector of length 3
\begin{eqnarray*}
\sum_{\mathbf{x}\in D_4}\frac{1}{\left(a+\left\Vert \mathbf{x}\right\Vert ^{2}\right)^{s}}
&=&
\sum_{j=0}^{3} \frac{r_{D_4}(j)}{(a+j)^s}+
\sum_{k=0}^{+\infty}\frac{\left(-1\right)^{k}a^k}{k!}\frac{\Gamma(s+k)}
{\Gamma(s)}\left(-3\cdot4^{2-s-k} \right)\\
&=&
\frac{1}{a^s}+\frac{24}{(a+2)^s}
-3\cdot 4^{2-s} \left(1-\frac{a}{4}\right)^s.
\end{eqnarray*}
\end{IEEEproof}

The implications of the above computations for the error probability are summarized below for $\gamma_e>1/2$. Similar expressions can be obtained 
for $\gamma_e>1/4$ (or smaller values of $\gamma_e$). 

\begin{cor}\label{cor:pce}
Suppose $\gamma_e > 1/2$.
We have when using $\Lambda_e=\ZZ^4$ that
\[
\bar{P}_{c,e}\leq C_{\mathrm{MIMO}} \gamma_e^{-2n_e} \left( \gamma_e^{2(n_e+2)} +\frac{8}{(1+1/\gamma_e)^{2(n_e+2)}}-8\cdot 4^{1-2(n_e+2)} 
\left(1-\frac{1}{4\gamma_e}\right)^{2(n_e+2)} \right)
\]
while with $\Lambda_e=D_4$ 
\[
\bar{P}_{c,e}\leq C_{\mathrm{MIMO}}\gamma_e^{-2n_e} \left( \gamma_e^{2(n_e+2)}-
3\cdot 4^{2-2(n_e+2)} \left(1-\frac{1}{4\gamma_e}\right)^{2(n_e+2)} \right).
\]
\end{cor}

\begin{figure}[ht]
\noindent \begin{centering}
\includegraphics[width=12cm]{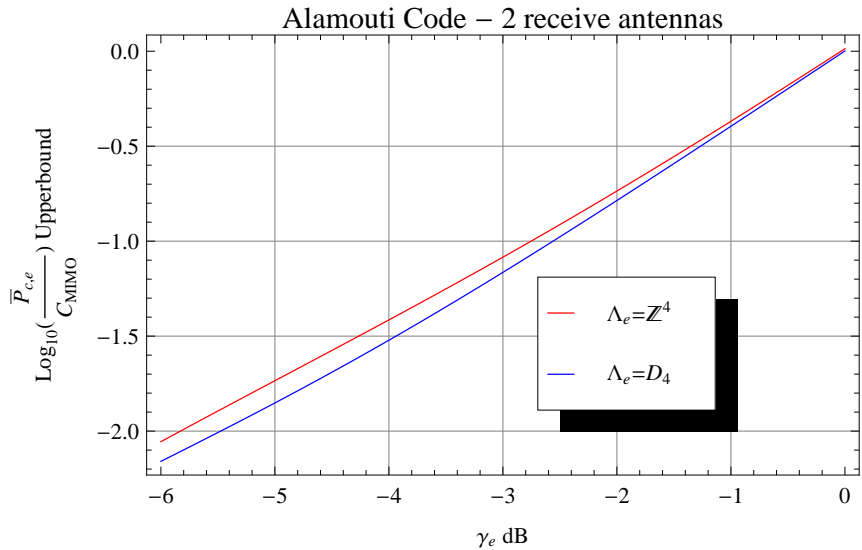}
\par\end{centering}
\caption{\label{fig:true-Alamouti} A tighter upper bound on $\tfrac{\bar{P}_{c,e}}{C_{\mathrm{MIMO}}}$: the Alamouti
code with $n_{e}=2$.}
\end{figure}

See Figure \ref{fig:true-Alamouti} for an illustration of the new bounds.

\begin{figure}[ht]
\noindent \begin{centering}
\includegraphics[width=12cm]{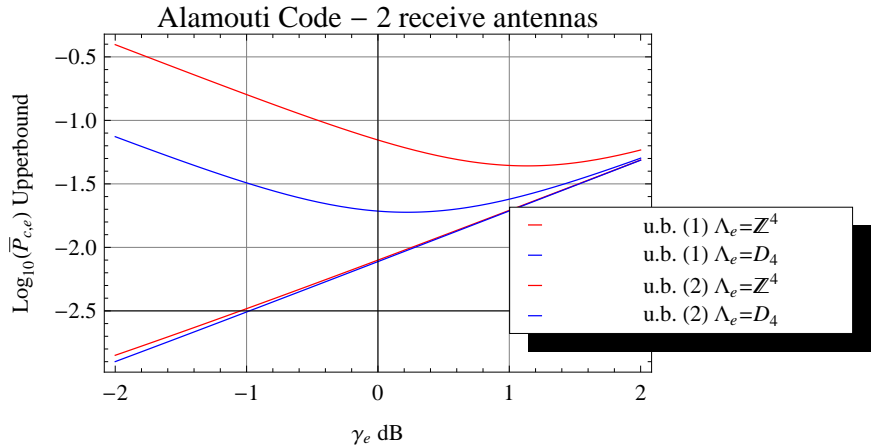}
\par\end{centering}
\caption{\label{fig:compare} Loose upper bounds versus tight upper bounds: the Alamouti code with $n_e=2$.}
\end{figure}

\begin{figure}[ht]
\noindent \begin{centering}
\includegraphics[width=12cm]{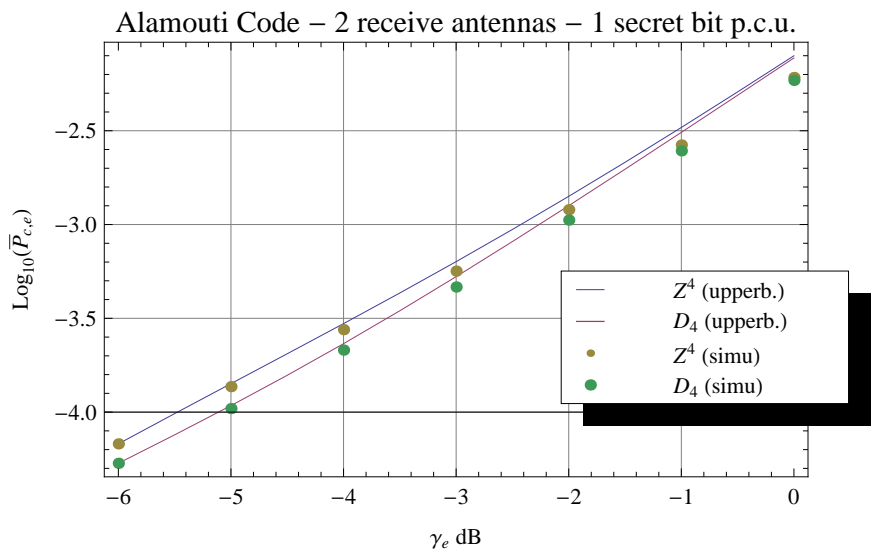}
\par\end{centering}
\caption{\label{fig:simul-Alamouti} Upper bounds versus simulations: the Alamouti code with $n_e=2$.}
\end{figure}

\textcolor{black}{
To conclude, we compare the loose upperbounds with the tight ones in Figure \ref{fig:compare}, and 
our bounds on the probability of correct decision for the eavesdropper with simulations 
in Figure \ref{fig:simul-Alamouti}. The coarse lattice $\Lambda_e$ is $\ZZ^4$ (resp. $D_4$) while the fine lattice
$\Lambda_b$ is $1/2 \ZZ^4$ (resp. $1/2 D_4$) giving rise to a secret spectral efficiency equal to 1 bit per real dimension.
For simulations, we used the linear ML decoder of the original Alamouti paper \cite{Alamouti}.
Decoding of $D_4$ has been done using the Wagner decoder of the binary parity check (4,3) code. 
}

\section{Conclusion}

We considered a MIMO wiretap channel, where Alice uses lattice codes via coset encoding to communicate with Bob in the presence of an eavesdropper Eve. We showed, by analyzing Eve's probability of correctly decoding the message meant to Bob, that this probability can be minimized by designing the lattice codes according to a suitable design criterion. The cases of block and fast fading channels are treated similarly. We also illustrate how our analysis applies to the Alamouti code, making explicit an interesting connection to Epstein zeta functions. 
Current and future work involve a more systematic design of such lattice wiretap codes.


\section*{Acknowledgments}

Part of this work was done while J.-C. Belfiore was visiting Nanyang Technological University, Singapore. The work of F. Oggier is supported by the Singapore National
Research Foundation under Research Grant NRF-RF2009-07.
The authors would like to thank Patrick Sol\'e for his help regarding Epstein zeta functions.

%
%

\appendix

In this appendix, we compute the Epstein zeta functions of $\ZZ^4$ and $D_4$.

Recall that the Epstein zeta function of a lattice $\Lambda$ is defined by
\begin{equation}
\boxed{\zeta_{\Lambda}(s)\triangleq\sum_{\mathbf{x}\in\Lambda\setminus\left\{ \mathbf{0}\right\} }\frac{1}{\left\Vert \mathbf{x}\right\Vert ^{2s}}=\sum_{n>0}\frac{r_{\Lambda}(n)}{n^{s}}}\label{eq:expr-zeta}
\end{equation}
where $r_{\Lambda}(n)$ is the number of vectors of $\Lambda$ with
a squared Euclidean norm equal to $n$. Note that $r_{\Lambda}(n)$ similarly appears in the theta series of 
$\Lambda$:
\[
\Theta_{\Lambda}(q)=1+\sum_{\mathbf{x}\in\Lambda\setminus\left\{ \mathbf{0}\right\} }q^{\left\Vert \mathbf{x}\right\Vert ^{2}}=1+\sum_{n>0}r_{\Lambda}(n)q^{n}.
\]

\begin{prop}\label{prop:zeta-z4}
The Epstein zeta function of $\ZZ_4$ is 
\[
\zeta_{\mathbb{Z}^{4}}(s) =  8\left(1-4^{1-s}\right)\zeta(s)\zeta(s-1).
\]
\end{prop}
\begin{IEEEproof}
We have 
\[
\zeta_{\mathbb{Z}^{4}}(s)=\sum_{n>0}\frac{r_{4}(n)}{n^{s}}
\]
where $r_{N}(n)$ is the number of solutions to the Diophantine equation
$k_{1}^{2}+k_{2}^{2}+\cdot\cdot\cdot+k_{N}^{2}=n$ (counting permutations
and signs). We now use a result of \cite[Paragraph 91]{Rademacher}, better
exposed in \cite{Borwein}, 
\[
r_{4}(n)=8\sigma(n)-32\sigma\left(\frac{n}{4}\right)
\]
where $\sigma(n)=\sum_{d|n}d$ and it is understood that $\sigma(m)=0$
if $m$ is not a positive integer. In particular, this implies that
\[
\sum_{n>0}\frac{\sigma(n/4)}{n^{s}}=\sum_{m>0}\frac{\sigma(m)}{(4m)^{s}}.
\]
We thus obtain
\begin{eqnarray}
\zeta_{\mathbb{Z}^{4}}(s) & = & 8\sum_{n>0}\frac{\sigma(n)}{n^{s}}-\frac{32}{4^{s}}\sum_{n>0}\frac{\sigma(n)}{n^{s}}\nonumber \\
 & = & 8\left(1-4^{1-s}\right)\sum_{n>0}\frac{\sigma(n)}{n^{s}}\nonumber \\
 & = & 8\left(1-4^{1-s}\right)\zeta(s)\zeta(s-1)
\end{eqnarray}
where the last equality comes from \cite[Chapter XVII]{hardy-wright}
and $\zeta(s)=\sum_{n>0}\frac{1}{n^{s}}$ is the Riemann zeta function. 
\end{IEEEproof}

\begin{prop}\label{prop:zeta-d4}
The Epstein zeta function of $D_4$ is 
\[
\zeta_{D_{4}}(s)  =  3\cdot4^{2-s}\left(2^{s-1}-1\right)\zeta(s)\zeta(s-1).
\]
\end{prop}
\begin{IEEEproof}
The lattice $D_{4}$ is the $4-$dimensional checkerboard lattice i.e., the set
of all $4$ dimensional integer valued vectors whose components have
an even sum. Its theta series is well-known \cite{CS-98} and is equal to
\[
\Theta_{D_{4}}\left(q\right)=\frac{1}{2}\left(\vartheta_{3}^{4}(q)+\vartheta_{4}^{4}(q)\right)
\]
where 
\begin{eqnarray*}
 \vartheta_3(q)^4 &=& (\sum_{k\in\ZZ}q^{k^2})^4\\
                  &=&  1+ \sum_{n>0}r_{4}(n)q^{n}
\end{eqnarray*}
is the theta series of $\mathbb{Z}^{4}$, whereas 
\begin{eqnarray*}
 \vartheta_4(q)^4 &=& (\sum_{k\in\ZZ}(-1)^kq^{k^2})^4 \\
                  &=& \sum_{k_1\in\ZZ}(-1)^k_1q^{k_1^2}\sum_{k_2\in\ZZ}(-1)^k_2q^{k_2^2}\sum_{k_3\in\ZZ}(-1)^k_3q^{k_3^2}\sum_{k_4\in\ZZ}(-1)^k_4q^{k_4^2}\\
                  &=&\sum_{k_1,k_2,k_3,k_4\in\ZZ}(-1)^{k_1+k_2+k_3+k_4}q^{k_1^2+k_2^2+k_3^2+k_4^2}\\
                  &=& 1+\sum_{n>0}(-1)^{n}r_{4}(n)q^{n}
\end{eqnarray*}
since $k_1+k_2+k_3+k_4 \equiv k_1^2+k_2^2+k_3^2+k_4^2 \mod 2$.
Thus the theta series of $D_4$ can be rewritten as
\begin{eqnarray*}
\Theta_{D_{4}}\left(q\right)&=&\frac{1}{2}\left( 1+ \sum_{n>0}r_{4}(n)q^{n}+ 1+\sum_{n>0}(-1)^{n}r_{4}(n)q^{n}\right)\\
&=& 1+ \frac{1}{2} \sum_{n>0} \left(  r_{4}(n)q^{n}+(-1)^{n}r_{4}(n)q^{n}\right),
\end{eqnarray*}
showing that the Epstein zeta function of $D_4$ is
\begin{eqnarray*}
\zeta_{D^{4}}(s) &=& \frac{1}{2} \sum_{n>0}\frac{ r_{4}(n)+(-1)^{n}r_{4}(n)}{n^{s}}\\
                 &=& \frac{1}{2} \zeta_{\ZZ^4}+\frac{1}{2} \sum_{n>0}\frac{(-1)^{n}r_{4}(n)}{n^{s}}.
\end{eqnarray*}
Using again as in the above proof that
\[
r_{4}(n)=8\sigma(n)-32\sigma\left(\frac{n}{4}\right),
\]
we get
\begin{eqnarray*}
\frac{1}{2}\sum_{n>0}(-1)^{n}\frac{r_{4}(n)}{n^{s}} & = & 4\left(\sum_{n>0}(-1)^{n}\frac{\sigma(n)}{n^{s}}-4\sum_{m>0}\frac{\sigma(m)}{\left(4m\right)^{s}}\right)\\
 & = & 4\left(\sum_{n>0}(-1)^{n}\frac{\sigma(n)}{n^{s}}-4^{1-s}\zeta(s)\zeta(s-1) \right)
\end{eqnarray*}
since \cite[Chapter XVII]{hardy-wright} 
\[
\sum_{n>0}\frac{\sigma(n)}{n^{s}}=\zeta(s)\zeta(s-1).
\]

We are left to compute 
\begin{eqnarray*}
\sum_{n>0}(-1)^{n}\frac{\sigma(n)}{n^{s}}
&=& -\sum_{n\,\mathrm{odd}}\frac{\sigma(n)}{n^{s}}+\sum_{n\,\mathrm{even}}\frac{\sigma(n)}{n^{s}}\\
&=&-\sum_{n\,\mathrm{odd}}\frac{\sigma(n)}{n^{s}}+\sum_{v>0}\frac{\sigma\left(2^{v}\right)}{2^{vs}}\sum_{n\,\mathrm{odd}}\frac{\sigma(n)}{n^{s}}\\
&=&\sum_{n\,\mathrm{odd}}\frac{\sigma(n)}{n^{s}}
\left(-1+  \sum_{v>0}\frac{2^{v+1}-1}{2^{vs}} \right)\\
&=&\sum_{n\,\mathrm{odd}}\frac{\sigma(n)}{n^{s}}
\left(-1+  \frac{2^{2-s}}{1-2^{1-s}}-\frac{2^{-s}}{1-2^{-s}} \right)
\end{eqnarray*}
where the second equality follows from the multiplicativity of $\sigma$.
Since similarly
\[
\sum_{n>0}\frac{\sigma(n)}{n^{s}}=\sum_{n\,\mathrm{odd}}\frac{\sigma(n)}{n^{s}}
\left(1+  \frac{2^{2-s}}{1-2^{1-s}}-\frac{2^{-s}}{1-2^{-s}} \right),
\]
a comparison of both expressions yields
\begin{eqnarray*}
\sum_{n>0}(-1)^{n}\frac{\sigma(n)}{n^{s}} &=&
\sum_{n>0}\frac{\sigma(n)}{n^{s}}\left(
\frac{-1+  \frac{2^{2-s}}{1-2^{1-s}}-\frac{2^{-s}}{1-2^{-s}}}{1+  \frac{2^{2-s}}{1-2^{1-s}}-\frac{2^{-s}}{1-2^{-s}}}
\right)\\
&=&
\zeta(s)\zeta(s-1)
\left(
\frac{\frac{2^{2-s}}{1-2^{1-s}}-\frac{1}{1-2^{-s}}}{\frac{2^{2-s}}{1-2^{1-s}}+\frac{1-2^{1-s}}{1-2^{-s}}}
\right)\\
&=&
\zeta(s)\zeta(s-1)
\left(
\frac{2^{2-s}(1-2^{-s})-(1-2^{1-s})}{2^{2-s}(1-2^{-s})+(1-2^{1-s})^2}
\right)\\
&=&
\zeta(s)\zeta(s-1)
\left(
2^{2-s}-2^{2-2s}+2^{1-s}-1
\right),
\end{eqnarray*}
and we obtain that
\begin{eqnarray*}
\frac{1}{2}\sum_{n>0}(-1)^{n}\frac{r_{4}(n)}{n^{s}} 
& = & 4\left(2^{2-s}-2^{2-2s}+2^{1-s}-1 -4^{1-s}\right) \zeta(s)\zeta(s-1)\\
&= &  4\left(2\cdot 2^{1-s}-2\cdot2^{2-2s}+2^{1-s}-1\right) \zeta(s)\zeta(s-1).
\end{eqnarray*}

We finally obtain the expression of the Epstein zeta function of $D_{4}$:
\begin{eqnarray*}
\zeta_{D_{4}}(s) & = &
\frac{1}{2} \zeta_{\ZZ^4}+\frac{1}{2} \sum_{n>0}\frac{(-1)^{n}r_{4}(n)}{n^{s}}\\
&=& 
4\left(1-4^{1-s}\right)\zeta(s)\zeta(s-1)+4\left(3\cdot 2^{1-s}-2\cdot2^{2-2s}-1\right) \zeta(s)\zeta(s-1)\\
&=&
4\left(-3\cdot 2^{2-2s}+3\cdot 2^{1-s}\right) \zeta(s)\zeta(s-1)\\
& = & 3\cdot4^{2-s}\left(2^{s-1}-1\right)\zeta(s)\zeta(s-1).
\end{eqnarray*}
\end{IEEEproof}


\end{document}